# Relaxor ferroelectricity and the freezing of short-range polar order in magnetite

F. Schrettle,[1] S. Krohns,[1] P. Lunkenheimer,[1,*] V. A. M. Brabers,[2] and A. Loidl[1]

[1]*Experimental Physics V, Center for Electronic Correlations and Magnetism, University of Augsburg, D-86135 Augsburg, Germany*

[2]*Department of Physics, Eindhoven University of Technology, NL-5600, MB Eindhoven, The Netherlands*

A thorough investigation of single crystalline magnetite using broadband dielectric spectroscopy and other methods provides evidence for relaxor-like polar order in $Fe_3O_4$. We find long-range ferroelectric order to be impeded by the continuous freezing of polar degrees of freedom and the formation of a tunneling-dominated glasslike state at low temperatures. This also explains the lack of clear evidence for a non-centrosymmetric crystal structure below the Verwey transition. Within the framework of recent models assuming an intimate relation of charge and polar order, the charge order, too, can be speculated to be of short-range type only and to be dominated by tunneling at low temperatures.

PACS numbers: 75.85.+t, 77.22.Gm, 77.80.Jk

# I. INTRODUCTION

Since Thales of Miletus studied magnetite more than 2500 years ago [1], this material has startled countless scientists. Its current applications range from compass needles over magnetic recording devices to ferrofluids and it is an important ingredient of many iron ores. In addition, several organisms like magnetotactic bacteria and pigeons use small magnetite crystals for orientation and the existence of such crystals in a meteorite was taken as evidence for former life on mars [2]. The prominent role of magnetite in solid state physics originates from the occurrence of a well-pronounced metal-insulator transition at $T_V \approx 122$ K, where the resistivity abruptly increases by two to three orders of magnitude. It is named after Verwey, who suggested charge ordering of $Fe^{2+}$ and $Fe^{3+}$ ions as its driving force [3]. However, more than 60 years of theoretical and experimental effort later, the nature of the Verwey transition still is under heavy debate [4].

At room temperature magnetite exhibits the cubic inverse spinel structure. The iron ions occupy interstices between $O^{2-}$ ions with $Fe^{3+}$ located at the tetrahedral sites, while the octahedral sites contain an equal amount of statistically distributed $Fe^{3+}$ and $Fe^{2+}$ ions. Below $T_V$ monoclinic and triclinic symmetries have been reported (e.g., [5,6,7,8]) and, despite many efforts, the low temperature structure of magnetite can be still regarded as unresolved. Among the vast literature on magnetite, dealing mostly with its magnetic and structural properties, during the last 35 years there were also some reports on magnetoelectric effects and electrical polarization in this material (e.g., [9,10]). Most of them were primarily intended to provide information on the controversially discussed crystal structure below $T_V$. However, in light of the recent tremendous interest in multiferroics, these reports have attained additional significance as they indicate that magnetite in fact may be the first known multiferroic compound. The most prominent case of multiferroicity is the coexistence of ferroelectricity and (anti)ferromagnetism [11]. Usually polar order as observed, e.g., in $BaTiO_3$ requires $d^0$-ness [12] and therefore ferroelectric (FE) ferromagnets are rare. In 2004 a new microscopic mechanism involving charge order (CO) driven ferroelectricity was proposed [13], which is assumed to explain the multiferroicity in $(PrCa)MnO_3$ [14] and $LuFe_2O_4$ [15]. Recently this model was suggested to also apply to the case of magnetite [16]. Very recent first-principles studies [6] and experimental results on thin films [17] seem to support this view.

However, from an experimental point of view, the appearance of ferroelectricity in magnetite is far from being settled. So far it is even unclear if the low-temperature structure allows for polar order. For example, high-resolution x-ray diffraction data point to centrosymmetric monoclinic symmetry [8], incompatible with ferroelectricity. In addition, to our knowledge FE hysteresis curves have been detected in two cases only, the results strongly deviating from those of canonical ferroelectrics [10,17]. Moreover, in semiconducting materials with the tendency of Schottky-diode induced surface-layer formation, the observation of FE hystereses has to be critically inspected [18,19]. To ensure an intrinsic origin, the temperature and frequency of the FE hysteresis experiments should be chosen on the basis of the temperature and frequency dependence of the complex dielectric constant $\varepsilon'$ [19]. Finally and most importantly, a conclusive investigation of magnetite by dielectric spectroscopy, demonstrating the typical characteristics of ferroelectrics, still is missing. Some dielectric anomalies involving high values of $\varepsilon'$ were observed at audio- or radio-wave frequencies, which, however, are strongly frequency dependent [17,20,21]. In contrast, high-frequency dielectric experiments in magnetite have revealed much smaller values of $\varepsilon'$ and no [22] or rather tiny anomalies, only [23]. It is well known, that transition metal oxides often exhibit non-intrinsic Maxwell-Wagner effects, with strongly enhanced values of $\varepsilon'$ and pronounced relaxation phenomena at low frequencies, which are difficult to distinguish from intrinsic contributions [24]. In the present work we provide a detailed investigation of the low-temperature dielectric properties of magnetite with broadband dielectric spectroscopy from mHz to GHz, aiming at a final clarification of the intrinsic dielectric properties and the occurrence of ferroelectricity in this compound.

## II EXPERIMENTAL DETAILS

A single crystal of magnetite was prepared from α-$Fe_2O_3$ using a floating-zone method with radiation heating [25]. Measurements of the specific heat between 2 and 300 K were performed utilizing a Quantum Design physical properties measurement system (PPMS). For the dielectric measurements, contacts of silver-paint or sputtered gold were applied at opposite sides of the platelike crystal. The dielectric properties at frequencies from 10 mHz to 1 MHz were determined using a frequency-response analyzer (Novocontrol α-Analyzer). Measurements between 1 MHz and 3 GHz were performed with a coaxial reflection technique, employing an impedance analyzer (Agilent E4991A) [26]. Applied ac voltages in these experiments were 1 V (low frequencies) and 0.4 V (high frequencies). FE hysteresis loops were measured with a Ferroelectric Analyzer (aixACCT TF2000) applying voltages up to 1.2 kV. For sample cooling down to 2 K, the Quantum Design PPMS and a closed-cycle refrigerator were used.

## III. RESULTS AND DISCUSSION

Figure 1 shows the temperature dependence of the heat capacity of magnetite on a double-logarithmic scale at 2 K < $T$ < 300 K. There are no indications of any further phase transition below $T_V$. Our results compare nicely with those in Ref. [27] but the anomaly at $T_V$ = 122.3 K is very narrow (width at half maximum ~ 0.8 K) and reaches values of more than 1000 J/(mol K). The released entropy is 8.9 J/(mol K), significantly larger than the values from literature, reported in [27]. These facts point towards superior sample quality. Below about 10 K, $C_p(T)$ shows a weaker increase than the $T^3$ power law expected for a purely harmonic solid and a stronger increase at about 10 - 30 K. This becomes obvious in the inset of Fig. 1 where the low temperature data are plotted as $C_p/T^3$ vs. $T$, which should lead to a constant behavior for pure phonon contributions. The deviations at low temperatures can be ascribed to magnon contributions, leading to a $T^{3/2}$ power law. The data below approximately 8 K indeed can well be fitted using $C_p = a\,T^{3/2} + b\,T^3$ (lines). Of high interest is the bumplike maximum close to 30 K revealed by the inset. It strongly reminds of the typical excess heat capacity of disordered solids and glasses. This is the first hint towards significant disorder in magnetite, still present at low temperatures.

Figure 2 shows $\varepsilon'(T)$ at $T < T_V$ for frequencies between $10^{-2}$ and $3\times10^9$ Hz (symbols). For the higher frequencies, $\varepsilon'(T)$ shows a steplike decrease from the static dielectric constant $\varepsilon_s$ of the order of several 1000 to its high-frequency limit $\varepsilon_\infty \approx 60$ when the temperature is lowered. This feature shifts to lower temperatures with decreasing frequency. These are the typical signatures of a dielectric relaxation process. Interestingly, at the lower frequencies, up to three such relaxation steps seem to be superimposed to each other. For ex-ample at 149 Hz, their points of inflection are approximately located at 10, 30, and 50 K. For $\nu \leq 16$ Hz, $\varepsilon'(T)$ shows a broad maximum below 40 K. With decreasing frequency, the height of the peak increases, while it is shifted to lower temperatures. This is the typical behavior of relaxor ferroelectrics, which are characterized by a diffuse phase transition and the freezing in of short-range clusterlike FE order [28,29]. It corresponds to a relaxation process with a strongly increasing $\varepsilon_s(T)$ as indicated by the high-temperature enveloping curve of the peaks (dashed line in Fig. 2), which was calculated assuming a Curie-Weiss law $C/(T-T_C)$ with C the Curie constant and $T_C$ the Curie temperature.

The question arises, which, if any, of the observed relaxations are of intrinsic nature. At first, it should be noted that the results at $\nu \leq 117$ kHz and $\nu \geq 1$ MHz were obtained during different measurement runs, and that the low-frequency measurements were performed with renewed contacts. Obviously, the values of $\varepsilon_s$ from both runs are different (Fig. 2), already pointing to non-intrinsic effects. Indeed, such high values of $\varepsilon'$ are often caused by electrode polarization arising from diode formation at the contact/sample interface [24]. For a further check, the low-frequency measurements were repeated with sputtered gold contacts instead of silver paint (solid lines in Fig. 2), which should affect the contact dominated regime only. Both, the height and the position of the major relaxation steps are strongly influenced. However, in the temperature range of the relaxor peak and the low-temperature relaxation, the second measurement is in line with the first one. In addition, the relaxorlike relaxation is consistent with the results reported in [21]. However, the lower frequencies investigated in the present work reveal a significantly shallower increase of $\varepsilon_s(T)$ (dashed line) than assumed in [21]. Also the relaxation found in ceramic samples in [20] seems to be consistent with the present results. We take all these findings as clear indication for the intrinsic nature of the observed relaxation dynamics below 50 K.

Even relaxor ferroelectrics should show well developed FE hystereses at low temperatures. Based on the results of Fig. 2, it is clear that hysteresis loops measured at sufficiently low temperatures and high frequencies should mirror the intrinsic behavior of magnetite. Figure 3 shows $P(E)$, obtained at 513 Hz and three temperatures. The curve at 5.6 K reveals a characteristic hysteresis loop that indicates FE behavior. In relaxor ferroelectrics such relatively broad loops are found at low temperatures, while the hystereses should narrow at higher temperatures [28,30]. However, in Fig. 3 the results at the higher temperatures are hampered by combined contributions from the strongly increasing conductivity in magnetite and the highly nonlinear contact effects (cf . Fig. 2) [18,19], which impedes a correction of the data. The FE polarization in magnetite, observed at 5.6 K, is by an order of magnitude lower than that in conventional ferroelectrics (at room temperature) and of the order of the spontaneous polarization of the spin-driven multiferroic $TbMnO_3$ [31]. The values for thin films, however, are by a factor of 10 larger [17], which may be ascribed to stress and strain effects in the thin films.

For a sound analysis of relaxational processes, the frequency dependence of $\varepsilon'$ and of the loss $\varepsilon''$ has to be considered. Figure 4 provides such plots for temperatures where, according to the above discussion, the dielectric response should be largely dominated by intrinsic contributions. The loss spectra of Fig. 4(b) were obtained after subtraction of the dc conductivity $\sigma_{dc}$, contributing to the loss via $\varepsilon'' \propto \sigma_{dc}/\nu$. The data show the typical signature of relaxational processes, namely a steplike decrease in $\varepsilon'(\nu)$ and a concomitant peak in $\varepsilon''(\nu)$. The peak frequency $\nu_p$ provides an estimate of the relaxation time $\tau \approx 1/(2\pi\nu_p)$. With decreasing temperature, both features continuously shift to lower frequencies by several decades, evidencing the freezing in of polar dynamics. Notably, the peaks are significantly broader than for the Debye case (half width >1.14 decades), which is a hallmark feature of glassy dynamics, ascribed to a disorder-induced distribution of relaxation times [32]. For temperatures below 24 K, the spectra reveal additional shoulders at around $10^3$ Hz, which are due to the weaker low-temperature relaxation discussed earlier. To obtain quantitative information on $\tau$, both $\varepsilon'(\nu)$ and $\varepsilon''(\nu)$ were simultaneously fitted using the empirical Havriliak-Negami function, $\varepsilon' - i\varepsilon'' = \varepsilon_\infty + (\varepsilon_s - \varepsilon_\infty)/\left[1 + (i\omega\tau)^{1-\alpha}\right]^\beta$ [33], commonly employed to describe relaxations in glassy matter [32]. $\alpha$ and $\beta$ determine the broadening and asymmetry of the curves. As shown by the lines in Fig. 4, reasonable fits of the main relaxation features are achieved in this way (deviations arise from the low-temperature relaxation).

As revealed by the inset of Fig. 4, the resulting $\tau(T)$ significantly deviates from Arrhenius behavior, $\tau = \tau_0 \exp[E/(k_B T)]$. When linearly extrapolating the high-temperature data points towards $1/T = 0$ (dashed line), we arrive at an unreasonably low attempt frequency $\nu_0 = 1/(2\pi\tau_0)$ of 15 MHz. The dash-dotted line in the inset of Fig. 4 indicates an Arrhenius law with a more realistic $\nu_0$ of 5 THz. Obviously, when approaching low temperatures the temperature dependence of the relaxation dynamics of magnetite becomes successively weaker. This indicates that it is increasingly dominated by temperature independent tunneling rather than by thermally activated processes.

## IV. SUMMARY AND CONCLUSIONS

Thus, overall magnetite reveals the typical signature of relaxor ferroelectricity and a continuous slowing down of its polar dynamics, dominated by tunneling at low tempera-tures. Our results resolve the never settled dispute about the low-temperature crystal symmetry of magnetite: For clusterlike ferroelectricity, the loss of inversion symmetry occurs on a local scale only and overall monoclinic symmetry can be retained [8]. However, the investigations revealing a non-centrosymmetric structure (e.g., [6,34]) may be sensitive to the local structure of the relaxor state in magnetite. Our findings also have implications for the CO in magnetite: Its polar and charge degrees of freedom should be intimately related [16] and, thus, also the CO should be of short-range type only, becoming frozen at low temperatures but never reaching a long-range ordered state. Instead, finally a slowly fluctuating glasslike state with $\tau > 100$ s is formed (a "charge glass"), dominated by quantum-mechanical tunneling. This scenario is supported by the observation of significant excess heat capacity, as routinely found in disordered matter (Fig. 1). The suppression of long-range CO in magnetite seems reasonable as the B-site ions form a strongly frustrated pyrochlore lattice. Already in the Anderson model of B-site ordering in ferrites [35] nearest-neighbor forces were shown to drive short range CO only and even long-range Coulomb interactions were found quite ineffective in inducing long-range order. In addition, the experimental evidence for long-range CO is far from being conclusive [4].

Overall our results solve the important questions of the presence of ferroelectricity in magnetite and of the discrepancy between the theoretically expected ferroelectricity and the frequent finding of a centrosymmetric structure. In addition, taking into account the close relation of ferroelectricity and charge order, our results suggest that the charge order in magnetite should be of short-range nature only.

**ACKNOWLEDGMENT**

This work was supported by the Deutsche Forschungsgemeinschaft via the Transregional

**Figure captions:**

FIG. 1. (color online) Temperature dependence of the heat capacity of magnetite on a double logarithmic scale. In the inset the specific heat is plotted as $C_p/T^3$ vs. $T$. The lines in main frame and inset were calculated by $C_p = a\,T^{3/2} + b\,T^3$ with $a = 0.172$ mJ/(mol K$^{5/2}$) and $b = 9.25\times10^{-5}$ mJ/(mol K$^4$).

FIG. 2 (color online). $\varepsilon'(T)$ of magnetite for various frequencies obtained with silver-paint (symbols) and sputtered gold contacts (solid lines). The dashed line, calculated assuming a Curie-Weiss law, illustrates the temperature dependence of $\varepsilon_s$ of the main intrinsic relaxation.

FIG. 3. Ferroelectric polarization $P$ of Fe$_3$O$_4$ as a function of external electric field $E$ at 513 Hz and three temperatures.

FIG. 4 (color online). $\varepsilon'(\nu)$ (a) and $\varepsilon''(\nu)$ (b) of magnetite (the latter corrected for the dc conductivity) for several temperatures below $T_V$. The lines are fits with the Havriliak-Negami function. The inset shows the resulting $\tau(T)$ in Arrhenius representation. The lines demonstrate Arrhenius laws as explained in the text.

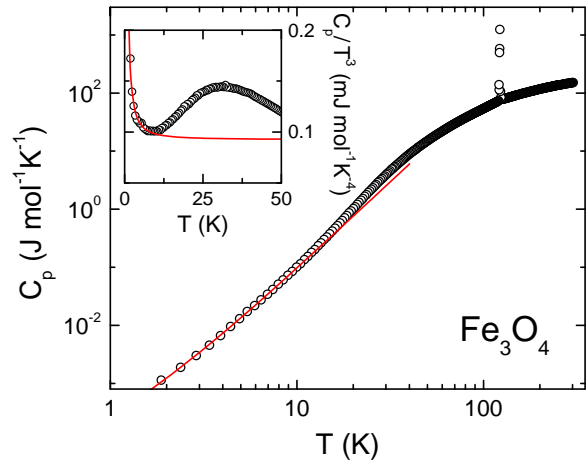

FIG. 1

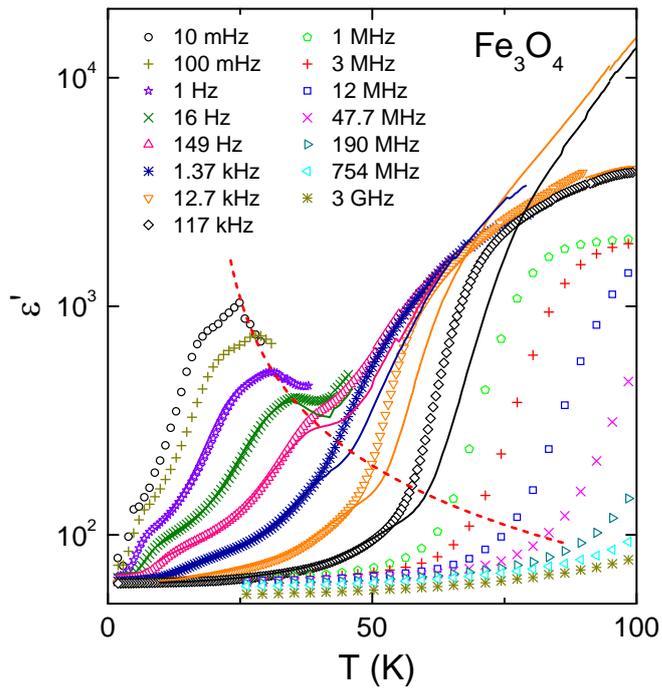

FIG. 2

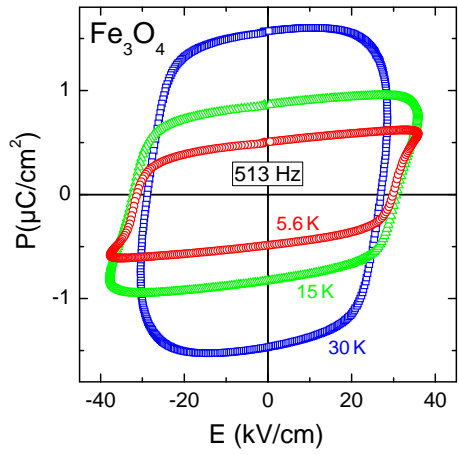

FIG. 3

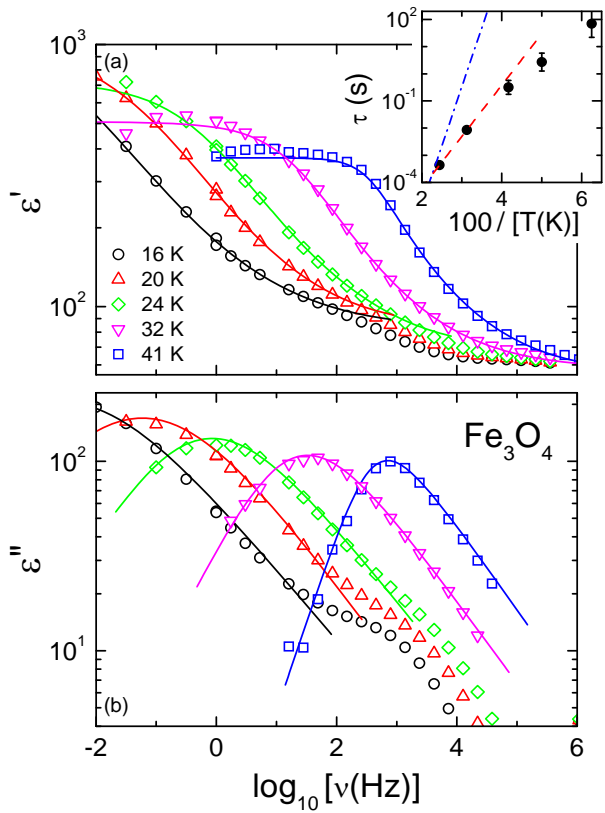

FIG. 4